\documentclass[runningheads]{llncs}
\usepackage[T1]{fontenc}
\usepackage{graphicx}
\usepackage{tikz}
\usetikzlibrary{tikzmark}
\tikzset{bullet/.style={circle,inner sep=0.2ex,fill}}
\usetikzlibrary{tikzmark,backgrounds} % add backgrounds library
\usetikzlibrary{matrix}
\usepackage{tabularx, makecell}
\usepackage{array}
\usepackage{colortbl}
\usepackage{lipsum}
\usepackage[hyphens]{url}
\usepackage{hyperref}
\usepackage{xcolor}
\usepackage{rotating}
\usepackage[table]{xcolor}  % add this in preamble
\usepackage{booktabs}
\usepackage{color}

\begin{document}
\title{Toward Self‑Organizing Production Logistics: A Multi-Agent Approach}
\titlerunning{Toward Self‑Organizing Production Logistics}
\author{Jan-Felix Klein\inst{1}\orcidID{0000-0002-3704-9567} \and
Yongkuk Jeong\inst{1}\orcidID{0000-0003-1878-773X} \and
Erik Flores-García\inst{1}\orcidID{0000-0003-0798-0753} \and Magnus Wiktorsson\inst{1}\orcidID{0000-0001-7935-8811}}
\authorrunning{J.-F. Klein et al.}
\institute{KTH Royal Institute of Technology, 10044 Stockholm, Sweden \\
\email{jfklein@kth.se}}
\maketitle              % typeset the header of the contribution
\begin{abstract}
Production logistics is increasingly exposed to variability, dynamic interdependencies, and operational disturbances that challenge conventional centralized planning and control approaches. Following a Design Science Research Methodology, this paper establishes a conceptual foundation for the design, implementation, and evaluation of Self-Organizing Production Logistics (SOPL) systems. First, key technological and systemic drivers motivating SOPL are identified, including autonomous logistics resources, advances in distributed AI-based decision-making, and the transition toward circular production systems, which further amplify operational uncertainty and complexity. Based on these drivers, system-level objectives and design requirements for SOPL are derived. Building on these requirements, the paper proposes an initial multi-agent architecture that integrates embodied and non-embodied agents, event-driven coordination, semantic knowledge structures, and digital twins. In addition, a three-phase demonstration roadmap is presented, progressing from an initial laboratory demonstrator toward increasingly distributed and adaptive SOPL systems. The Phase I demonstrator provides an experimental environment for investigating disturbance handling, human involvement, and supervisory coordination within an order-driven kitting and supply scenario.

% The abstract should briefly summarize the contents of the paper in

\end{abstract}
\section{Introduction}

\label{sec:introduction}
% Initial Motivation:
Production logistics (PL), also referred to as manufacturing logistics, encompasses storing, handling or transportation activities that support production planning, control, configuration and execution \cite{strandhagen_fit_2017}. 
Trends toward mass customization, personalization, and shorter product life cycles increase the number of part variants and smaller batch sizes \cite{hu_evolving_2013}, thereby increasing material flow complexity and material handling effort within PL systems. 
At the same time, flexible matrix and network-oriented production structures introduce dynamic routing decisions and stronger interdependencies between logistics resources and production processes \cite{greschke_matrix-produktion_2016}.
In parallel, circular production systems introduce amplified uncertainty through reverse flows of returned end-of-life products and components, whose quality, condition, and compatibility are often only revealed during inspection and execution \cite{fleischer_self-learning_2024}. This requires logistics and production decisions to be continuously adapted based on real-time system states. From a Complex Systems Theory perspective, such increasing system complexity leads to emergent and partially unpredictable behavior, thereby increasing the likelihood of disturbances during execution \cite{saurin_assessing_2013}. While these characteristics are particularly pronounced in circular production, similar trends toward higher variability, decentralization, and operational uncertainty are increasingly observed across modern production logistics systems. Circular factories are therefore used in this paper as a representative boundary case in which these dynamics become especially visible.

Traditionally, production planning and control have predominantly relied on centralized strategies that assume stable process behavior, high-quality system information, and the feasibility of computing globally optimal or near-optimal plans offline. While such approaches ensure stability and efficient resource utilization, they often lack responsiveness under dynamic disturbances and increasing variability \cite{diaz_c_non-centralised_2021,mes_comparison_2007}. Decentralized control approaches improve adaptability by distributing decision-making to local entities, but may result in globally suboptimal behavior due to limited coordination and information availability \cite{hulsmann_changing_2007}. Consequently, current research increasingly investigates hybrid architectures \cite{ismayyir_modelling_2024} that combine global coordination with localized autonomy.

Inspired by biological systems such as flocks of birds or schools of fish \cite{camazine_self-organization_2003}, the concept of self-organization has been increasingly adopted in logistics and manufacturing research (SOL, SOMS) \cite{gerrits_self-organizing_2023,qin_self-organizing_2021}. Self-organizing systems operate without centralized control, instead relying on decentralized decision-making and local interactions that give rise to emergent global behavior \cite{di_marzo_serugendo_self-organising_2011}. In production logistics, these principles translate into autonomous agents capable of local decision-making, negotiation, and adaptive coordination under real-time system conditions.

Despite growing interest, practical realization remains limited due to challenges in interoperable communication, adaptive control mechanisms, and scalable coordination architectures \cite{qin_self-organizing_2021}. At the same time, recent advances in foundation models enable new forms of adaptive, context-aware decision-making \cite{liu_llm-enhanced_2026}. Combined with embodied AI and semantic knowledge representations, these developments open new opportunities for implementing self-organizing principles through interacting multi-agent systems.
This paper is positioned as an initial conceptual step within a broader research agenda guided by the following research question:
\begin{quote}
    \textit{How can multi-agent AI systems be designed to realize Self-Organizing Production Logistics (SOPL) and thereby improve responsiveness and resilience under disturbances?}
\end{quote}
\vspace{5pt}
On the broader path toward answering this research question, this paper contributes by 
\begin{itemize}
    \item identifying key drivers of SOPL,
    \item proposing an initial multi-agent AI architecture, and
    \item outlining a three-phase demonstration roadmap toward experimental realization.
\end{itemize}

\section{Methodology}
\begin{figure} [t]
    \centering
    \includegraphics[width=1.0\linewidth]{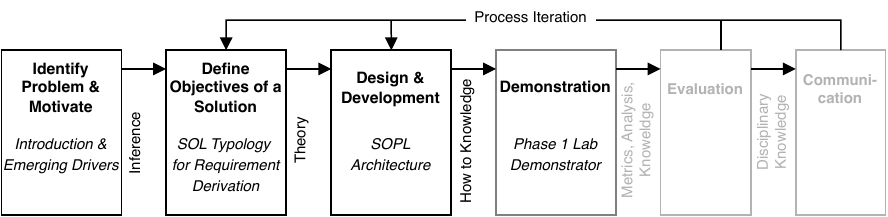}
    \caption{Applied Design Science Research Methodology (DSRM) process \cite{peffers_design_2007} for the development of SOPL systems.}
    \label{fig:methodology}
\end{figure}
The research follows the Design Science Research Methodology (DSRM) proposed by Peffers et al. \cite{peffers_design_2007}, see Figure \ref{fig:methodology}. The methodology is used as a structuring framework for the design and conceptual development of a multi-agent artifact for enabling SOPL. The work is positioned as an early-stage conceptual design study and primarily addresses the problem identification, objective definition, and design \& development phases of the DSRM cycle with initial prototypical demonstration activities considered in the roadmap.
Across all DSRM activities, the circular factory context \cite{lanza_vision_2024} is used as a representative high-variability production environment characterized by complex interdependent material flows, and frequent disturbances. It is therefore treated as a target environment for the development of SOPL systems, as it represents a setting in which traditional planning and control approaches face significant limitations.
Following the DSRM structure, the subsequent sections correspond to the main phases of the methodology. Section \ref{sec:drivers} continues the problem identification and motivation by analysing three recent drivers that further justify research into self-organizing systems. Section \ref{sec:objectives} derives the objectives for SOPL solutions, using the SOL typology as a structuring framework. Section \ref{sec:architecture} presents an initial architecture of the envisioned system, aligned with the circular factory environment. Finally, Section \ref{sec:roadmap} outlines an implementation roadmap, including an initial demonstrator lab scenario and directions for future iterations.

\section{Technological and Systemic Drivers of SOPL}
\label{sec:drivers}

Building on the increasing complexity and disturbance-prone nature of modern production logistics systems outlined in the introduction, this section identifies three key drivers for Self-Organizing Production Logistics (SOPL). Two drivers originate from technological advances that extend the autonomy and adaptability of physical logistics resources and AI-based decision-making systems. The third driver arises from the transition toward circular production systems, which introduces highly dynamic and uncertainty-rich operational conditions and therefore represents a particularly challenging application context for SOPL.

\subsection{Increasing Autonomy of Production Logistics Resources}
\label{sec:drivers_resources}

Production logistics resources are undergoing rapid transformation driven by Industry 4.0 technologies, including Cyber-Physical Production Systems (CPPS), the Industrial Internet of Things (IIoT), and data-driven infrastructures. Industrial deployment trends confirm this development, with strong growth in service robots for logistics and increasing adoption of autonomous mobile systems \cite{IFR2025ServiceRobots,LogisticsIQ2025AGVAMR}. Beyond traditional platforms, emerging robotic systems such as humanoids further expand the functional scope of logistics resources toward more general-purpose manipulation tasks \cite{eser_automation_2026}. As a result, logistics resources are increasingly evolving towards more autonomous and locally decision-capable actors, strengthening their role as active and adaptive elements within SOPL systems.

\subsection{Increasing AI Capabilities for Distributed Decision-Making}
\label{sec:drivers_ai}

Recent advances in artificial intelligence (AI) are enabling new forms of adaptive and distributed decision-making. Large language models (LLMs) provide general-purpose reasoning capabilities for interpreting instructions, integrating contextual information, and supporting decision-making across heterogeneous tasks and modalities \cite{raza_industrial_2025}. Building on these capabilities, LLM-based multi-agent systems are emerging as a promising approach for distributed reasoning and collaborative problem-solving \cite{tran_multi-agent_2025}. In such systems, specialized agents interact, exchange information, and jointly generate decisions under dynamic conditions.

In parallel, advances in multimodal reasoning enable the integration of textual, sensory, visual, and symbolic information into unified decision processes \cite{yang_magma_2025}. Furthermore, progress in knowledge integration and semantic reasoning allows these systems to incorporate explicit domain knowledge into decision-making \cite{pan_unifying_2024}. Together, these developments increase the feasibility of distributed, context-aware coordination mechanisms and thereby support the realization of self-organizing principles in production logistics systems.

\subsection{Circular Production as a Driver for SOPL}
\label{sec:circular_production}

Circular production systems require the tight integration of reverse flows for component recovery with forward flows for assembly and manufacturing. The circular factory concept exemplifies this development by enabling the concurrent production of current product generations while selectively reusing components recovered from returned cores of previous generations \cite{lanza_vision_2024}. As a result, individual product instances are increasingly characterized by unique combinations of reused and newly manufactured components.

These conditions introduce high variability in component quality and availability, uncertain return timings, cross-generational compatibility constraints, and batch sizes that frequently approach one. Consequently, production and logistics processes are increasingly affected by structural uncertainty, making long-term globally optimized planning approaches progressively less effective \cite{fleischer_self-learning_2024}. The circular factory therefore represents a particularly demanding and representative application environment for SOPL systems, requiring execution-close adaptation and real-time coordination.

\section{Objectives of an SOPL system}
\label{sec:objectives}
Building on the drivers outlined in the previous section, the objective of an SOPL system is not only to decentralize decision-making, but to enhance the ability of production logistics systems to operate effectively under structural uncertainty. Consequently, the objectives are formulated on two complementary levels: (i) system-level performance objectives that describe desired logistics outcomes, and (ii) design requirements that define the system properties required to achieve these outcomes.
The system-level performance objectives were derived by consolidating recurring performance expectations related to self-organizing systems and decentralized control in production environments \cite{bartholdi_self-organizing_2010,li_manufacturing_2025,gerrits_self-organizing_2023,qin_self-organizing_2021}. SOPL should improve responsiveness and resilience under uncertainty, enable scalable adaptability, and support continuous improvement, while safeguarding core logistics performance. These objectives reflect the need for production logistics systems that can react to changing conditions without compromising key targets such as throughput, service level, and operational stability.
To structure the corresponding design requirements, this paper adopts the Self-Organizing Logistics (SOL) typology \cite{gerrits_towards_2024} as an analytical lens. Its dimensions of system architecture, cooperativeness, autonomy, and system features provide a structured basis for identifying the forms of decentralization, interaction, and adaptation required for SOPL. From this perspective, decentralized coordination and learning-based adaptation emerge as central enablers of self-organizing behavior.
However, the SOL typology does not fully capture several requirements that are critical in production logistics contexts. These include modular and reconfigurable physical assets, shared semantic knowledge structures, and human governance mechanisms. These aspects are therefore introduced as complementary design requirements, extending the SOL perspective toward the specific needs of SOPL systems.

\begin{table}[t!]
\caption{Traceability matrix linking system-level performance objectives and design requirements.}
\centering
\setlength{\tabcolsep}{5.3pt}
\renewcommand{\arraystretch}{1.5}

\newcolumntype{L}[1]{>{\raggedright\arraybackslash}m{#1}}
\newcolumntype{C}[1]{>{\centering\arraybackslash}m{#1}}
% \fontsize{10pt}{10pt}\selectfont
\footnotesize
\begin{tabularx}{\textwidth}{
L{2.2cm}ccccc}
\toprule
\textbf{Performance objective} &
\makecell{Decentralized\\coordination} &
\makecell{Modular\\assets} &
\makecell{Semantic\\knowledge} &
\makecell{Learning \&\\adaptation} &
\makecell{Human\\governance} \\
\midrule
Improve responsiveness & ++ & + & + & + & o \\
Improve resilience & + & ++ & ++ & + & + \\
Enable scalable adaptability & ++ & ++ & ++ & + & o \\
Enable continuous improvement & + & o & + & ++ & + \\
Safeguard core logistics performance & o & + & ++ & + & ++ \\
\bottomrule
\end{tabularx}
% \caption{Traceability matrix linking system-level performance objectives and design requirements}
\label{tab:objective_matrix}
\end{table}

Table \ref{tab:objective_matrix} summarizes this relationship. The rows represent system-level performance objectives, while the columns represent key design requirements of SOPL systems. The symbols indicate the relative strength of contribution of each requirement to each objective, where ++ denotes a strong contribution, + a relevant contribution, and o a supporting contribution. The strength indicators are qualitative and represent the relative conceptual importance of a design requirement for achieving a given performance objective. 

Two insights emerge. First, responsiveness and scalable adaptability are strongly enabled by decentralized coordination and modular assets, as both support local decision-making and structural reconfiguration close to execution. Second, safeguarding core logistics performance depends particularly on shared semantic knowledge and human governance, which ensure global coherence and prevent local autonomy from degrading system-wide objectives. The matrix thus provides an analytical bridge between system requirements and the architectural design presented in the following section.

\section{Architecture for SOPL}
\label{sec:architecture}
\begin{figure}[t!]
    \centering
    \includegraphics[width=1.0\linewidth]{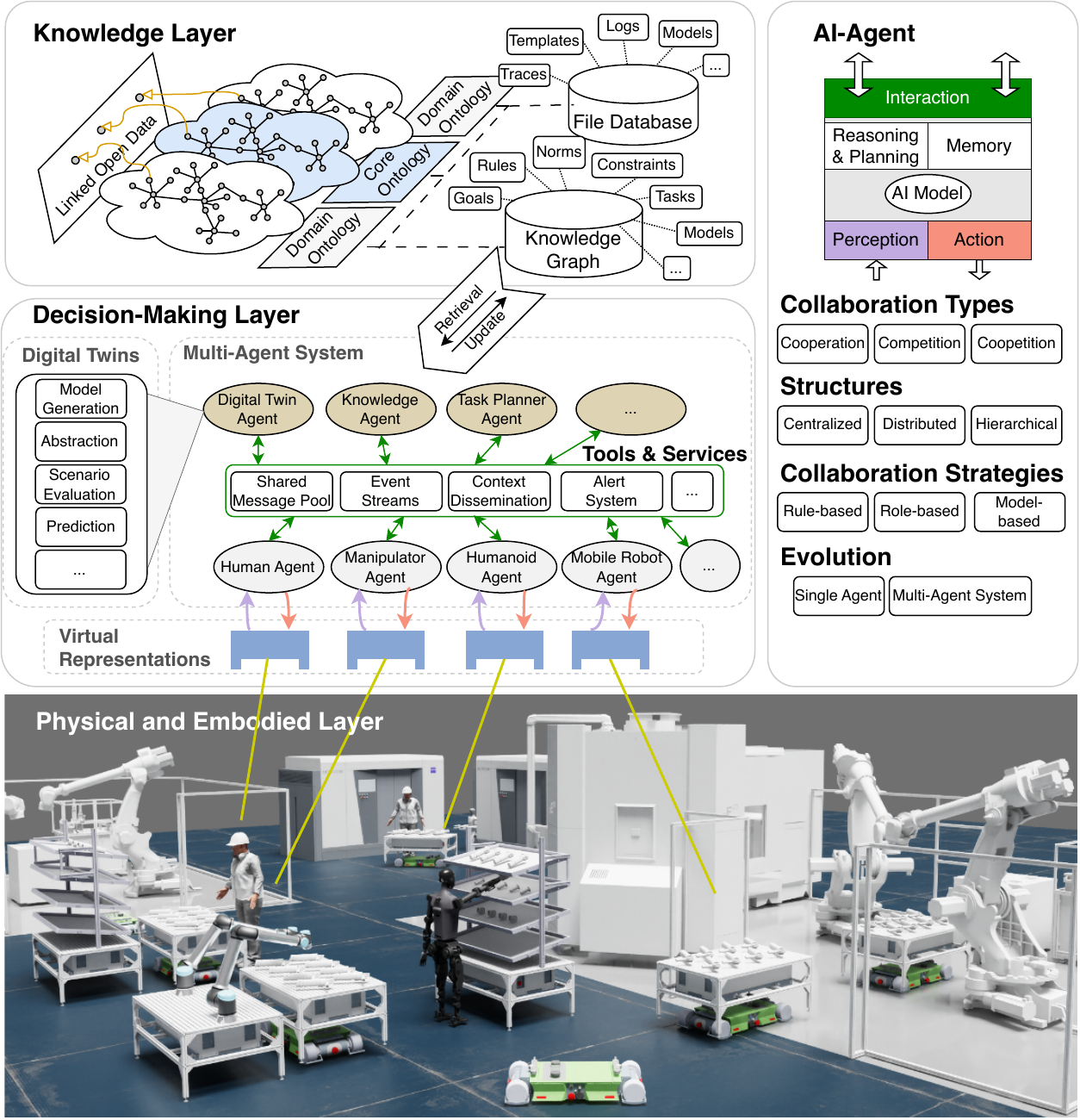}
    \caption{Initial SOPL architecture combining heterogeneous assets, agent-based decision-making, and shared knowledge. The depicted shop floor builds on modular intralogistics resources developed for the circular factory environment \cite{klein_knowledge-based_2025}, while the overall architecture draws on elements from \cite{li_survey_2024,tran_multi-agent_2025}.}
    \label{fig:vision}
\end{figure}

The objective matrix in Table \ref{tab:objective_matrix} indicates that the targeted system-level performance of SOPL depends on the interplay of several design requirements. Accordingly, the architectural challenge is to integrate the corresponding design elements into a coherent system. Figure \ref{fig:vision} illustrates an initial architecture for such an SOPL system. It is structured into three interrelated layers: the physical and embodied layer, the decision-making layer, and the knowledge layer The following subsections discuss each layer and its contribution to the identified design requirements.

\subsection{Physical and Embodied Layer}
\label{sec:embodied_layer}
The physical and embodied layer comprises the set of heterogeneous resources through which production logistics is executed. A core assumption of the proposed architecture is that future production logistics will rely on a diverse portfolio of assets rather than a single dominant automation paradigm. System capability is therefore not concentrated in one resource type but emerges from the combination of complementary assets that jointly cover transport, storage, buffering, handling, kitting, inspection support, and material presentation.

This asset portfolio includes conventional infrastructure such as conveyor systems and automated storage systems, which remain essential for stable, high-throughput flow. It further includes autonomous mobile robots, automated guided vehicles, cobots, and emerging embodied systems such as humanoids that provide flexible execution capabilities at the point of use. In addition, modular logistics units and process modules extend the capability space by enabling reconfiguration of functions such as buffering, kitting, or temporary handling. Depending on the application, such modules can also serve as interchangeable process components or task-specific extensions of robotic systems.

Modularity plays a central role, as it enables capability composition under resource constraints. Instead of relying on permanently dedicated infrastructure for each logistics function, modular assets can be combined, repositioned, and repurposed according to current system demands. This increases adaptability while maintaining efficient resource utilization, particularly under changing layouts, product structures, or process dependencies.

Within this layer, humans are explicitly modeled as high-capability and knowledge-intensive resources. They contribute contextual interpretation, exception handling, and decision-making under uncertainty, particularly in situations with incomplete or ambiguous information. The embodied layer therefore consists of a coordinated mix of conventional automation, modular systems, advanced robotic agents, and human actors, forming the physical basis for decentralized coordination.

\subsection{Decision-Making Layer}
\label{sec:decision_making}
The decision-making layer is realized through a distributed multi-agent system. An agent is defined as a computational entity situated within a production logistics environment and capable of autonomous action in pursuit of defined objectives \cite{wooldridge_introduction_2009}.
Two types of agents must be distinguished. Embodied agents are directly associated with physical resources. Their decisions are closely coupled to physical execution. Importantly, this coupling is not assumed to occur through direct and unrestricted access to the physical system. Instead, the agent interacts with the asset through a virtual representation that acts as an intermediate layer between decision-making and execution. This representation provides structured access to the physical resource, exposes relevant state and action information, and helps ensure safe and controlled interaction with the underlying asset. 
Non-embodied agents, in contrast, do not directly execute physical actions. They support coordination, reasoning, monitoring, simulation, and orchestration tasks. This includes functions such as task decomposition, system-level coordination, predictive analysis, and interaction with shared services.
Agents are endowed with a set of unique attributes and capabilities that define its behavior patterns and role within the environment \cite{li_survey_2024}. Coordination in this layer emerges from interaction among agents. Embodied agents contribute execution-relevant decisions such as accepting tasks, reporting state changes, or negotiating feasible allocations, while non-embodied agents can provide decision support, mediate information exchange, or maintain coordination structures. Communication is enabled through shared message spaces, event streams, and negotiation protocols.

Within the decision-making layer, digital twins (DTs) serve as dynamic, model-based decision-support instruments. On the basis of virtual models, they enable agents to anticipate consequences, compare alternatives, and assess decisions prior to physical execution. The proposed architecture assumes a specialized DT agent that coordinates simulation tasks across the relevant agents, parameterizes scenarios at an appropriate level of abstraction, and provides the resulting information in a form suitable for decentralized decision-making.

\subsection{Knowledge Layer}
The knowledge layer provides the shared semantic foundation required for coordination in highly variable production environments. Its importance is particularly evident in circular production settings, where products and components vary not only in type but also in condition, origin, lifecycle state, and compatibility. Under these conditions, consistent interpretation of system entities becomes a prerequisite for coordinated action.
To support this, the knowledge layer is based on ontologies capturing key domain concepts such as products, processes, resources, capabilities, tasks, and operations. These ontologies are instantiated in a shared knowledge graph and complemented by structured data repositories \cite{hofmann_role_2024}, enabling consistent access to both static and dynamic system information. Beyond semantic interoperability, the knowledge layer also supports governance and adaptation. It encodes constraints, rules, safety requirements, and escalation mechanisms that guide and limit autonomous behavior. At the same time, execution logs and outcome data are stored to enable continuous refinement of coordination policies, capability models, and system knowledge.
Overall, the knowledge layer acts as the semantic backbone of the architecture, ensuring coherent interpretation, explainability, and adaptive coordination across all agents and system components.

\section{Demonstration Roadmap}
\label{sec:roadmap}

\begin{figure} [!t]
    \centering
    \includegraphics[width=1.0\linewidth]{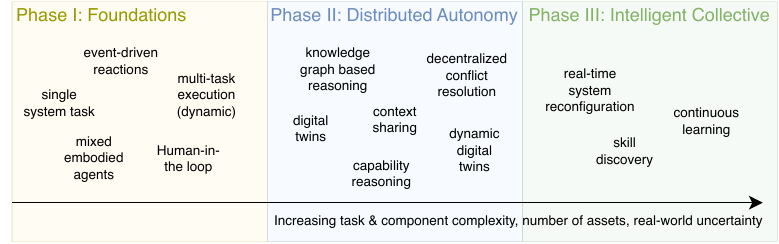}
    \caption{A three-phase demonstration roadmap toward SOPL.}
    \label{fig:roadmap}
\end{figure}
To align the conceptual architecture with the demonstration stage of the DSRM process, a three-phase demonstration roadmap is proposed, as illustrated in Figure \ref{fig:roadmap}. The roadmap starts with a \textit{foundation} (I) phase, in which a basic SOPL architecture is realized in a controlled laboratory environment. It then extends toward a \textit{distributed autonomy phase} (II), in which semantic reasoning, and more dynamic coordination mechanisms are progressively introduced. Finally, the roadmap points toward an \textit{intelligent-collective phase} (III), characterized by continuous learning, large-scale coordination, and operation under higher real-world variability. The following section focuses on the initial phase 1 demonstrator which establishes the baseline for subsequent development.

\subsection{Phase I: Foundations}
\label{sec:phase1}
The first phase realizes an initial SOPL demonstrator in a controlled laboratory environment. It serves primarily as a proof of concept for the technical feasibility of implementing heterogeneous embodied and non-embodied agents endowed with LLM-based reasoning capabilities, and for studying their coordination behavior in distributed task execution enabled by event-driven mechanisms and shared message pools.
The implementation is guided by the following research questions:
\begin{itemize}
\item \textbf{RQ1}: How do individual LLM-based agents react to disturbance alerts, and which initial decision strategies do they apply to restore task execution?
\item \textbf{RQ2}: How can human operators be integrated effectively into disturbance-handling processes?
\item \textbf{RQ3}: What role does a supervisor agent play in coordinating disturbance resolution across heterogeneous agents, and how does its involvement influence the quality of decision-making?
\end{itemize}

\begin{figure}[t!]
    \centering
    \includegraphics[width=1.\linewidth]{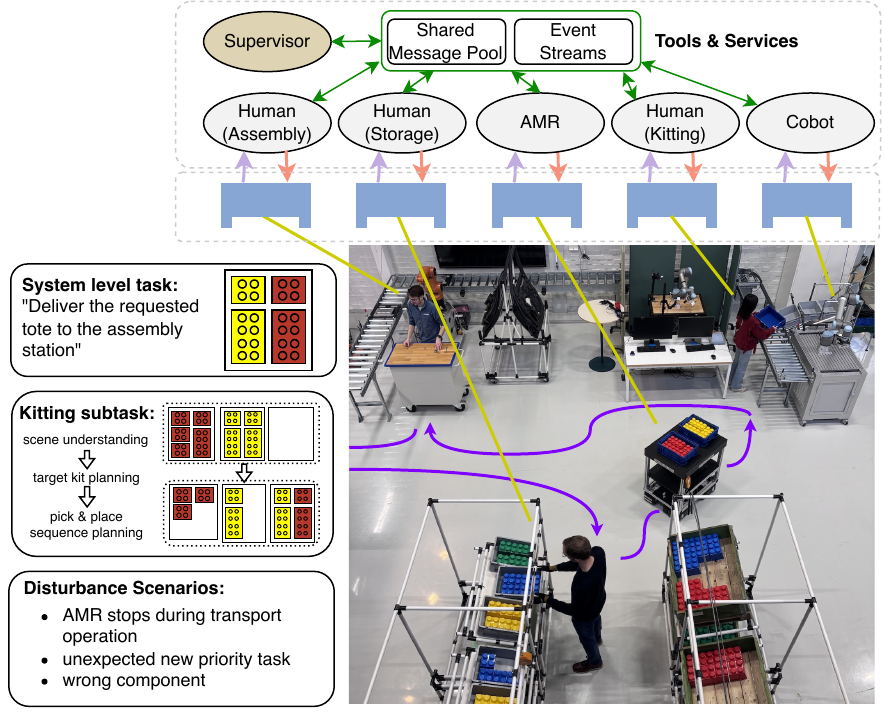}
    \caption{Phase 1 demonstration environment in the IPU Lab at KTH, where heterogeneous agents collaboratively perform an order-driven kitting and supply task.}
    \label{fig:phase_1}
\end{figure}

The demonstration scenario set in the IPU Lab at KTH, shown in Figure \ref{fig:phase_1}, is centered on an order-driven kitting and supply task in which an assembly station requests a kit of components. The fulfillment of this request comprises the retrieval of required components from storage, order picking, transport to a kitting station, assembly of the requested kit in a tote, and final delivery to the requesting assembly station.

The task is executed by a set of heterogeneous agents with complementary roles. The embodied agents represent the physical resources involved in the process, including an AMR responsible for transport operations and a picking cobot that performs the kitting subtask through scene understanding, target-kit planning, and pick-and-place sequence planning. The system further includes three human agents: one performing order picking in the storage area, one handling totes at the kitting station, and one receiving and confirming the requested totes at the assembly station.
In addition to these embodied actors, the demonstrator includes optional non-embodied software agents to support coordination and disturbance handling. Among these, a supervisor agent is treated as an experimental design parameter. Depending on the configuration, it may maintain a system-level view of task progress, monitor reported events and disturbances, and support the coordination of recovery actions across the involved agents. Communication between agents is realized through event-driven mechanisms and shared message pools, through which task requests, state changes, completion notifications, and disturbance alerts are exchanged.

To evaluate the disturbance-handling capabilities of the demonstrator, three predefined disturbance scenarios are introduced. First, a resource-related disturbance is considered in which the AMR stops during a transport operation, interrupting the physical flow of materials between process stages. Second, a process-related disturbance is introduced through the arrival of an unexpected new priority task, requiring the involved agents to reassess the current execution sequence and respond to changing operational priorities. Third, a product-related disturbance is examined in the form of a wrongly picked component at the storage area, creating a mismatch between the requested and the actually supplied material. Together, these scenarios provide a controlled basis for observing how individual agents interpret disturbance alerts, which initial response strategies they apply, how human operators are involved in recovery, and to what extent the optional supervisor agent contributes to the coordination of resolution actions.

\subsection{Phases II and III: Toward Distributed Autonomy and Intelligent Collectives}
\label{sec:phase2_3}
Beyond the initial demonstrator, the roadmap foresees a gradual transition toward more distributed autonomy and, ultimately, toward an intelligent collective capable of operating under the higher variability and uncertainty. In Phase II, task complexity, operational realism, and the number of interacting agents increase, moving the system beyond tightly structured single-task workflows toward more dynamic multi-task settings. To support this transition, agents are equipped with richer semantic descriptions of tasks, resources, and capabilities, enabling context-aware reasoning, capability-based task allocation, and more decentralized coordination through negotiation and conflict resolution. At the same time, digital twins as collaborative decision support tool are expected to be integrated, see Section \ref{sec:decision_making}. Building on these foundations, Phase III targets the realization of a continuously learning and highly adaptive SOPL system that can cope with heterogeneous product structures, fluctuating resource availability, uncertain component conditions, and recurring disturbances. In this stage, agents increasingly refine their decision policies and coordination strategies through historical interaction data, digital-twin traces, and operational feedback, enabling more proactive behavior, autonomous reorganization of responsibilities, and sustained alignment with system-level production objectives. Together, these phases describe the progression from a controlled proof of concept toward a knowledge-based, self-improving, and robust collective.

\section{Conclusion and Future Work}
\label{sec:conclusion}
This paper presented a DSRM-oriented process toward the development of Self-Organizing Production Logistics (SOPL) systems. Motivated by increasing variability, complexity, and disturbance proneness in future production-logistics environments, with circular production as one prominent example, it proposed a layered multi-agent architecture combining heterogeneous embodied and non-embodied agents, event-driven coordination, shared message pools, semantic knowledge structures, and digital twins. The paper further structured the path toward implementation through a three-phase roadmap, with particular focus on an initial demonstrator. In this way, it provides a conceptual foundation for the design and staged realization of SOPL systems.

Future work will focus first on the systematic design and evaluation of the Phase I demonstrator. In particular, a design of experiments is needed to define which disturbance scenarios, system configurations, and performance indicators should be assessed in order to answer the demonstrator research questions. This includes evaluating how agents respond to disturbances, how human operators are integrated into recovery processes, and how the presence of a supervisor agent affects coordination quality. A second priority is an industry study targeting disturbance-prone production-logistics scenarios in brownfield environments. The aim is to identify industrial cases that motivate SOPL, analyze the structural characteristics of existing systems, and derive requirements for their gradual transformation toward SOPL-compatible architectures.

\begin{credits}
\subsubsection{\ackname} 
This research was supported by the Centre of Excellence in Production Research (XPRES) as well as by the German Research Foundation (DFG) - SFB 1574 – 471687386

\subsubsection{\discintname}
The authors have no competing interests to declare that are relevant to
the content of this article.
%It is now necessary to declare any competing interests or to specifically
%state that the authors have no competing interests. Please place the
%statement with a bold run-in heading in small font size beneath the
%(optional) acknowledgments\footnote{If EquinOCS, our proceedings submission
%system, is used, then the disclaimer can be provided directly in the system.},
%for example: The authors have no competing interests to declare that are
%relevant to the content of this article. Or: Author A has received research
%grants from Company W. Author B has received a speaker honorarium from
%Company X and owns stock in Company Y. Author C is a member of committee Z.
\end{credits}

\bibliographystyle{splncs04}
\bibliography{references}



% LNCS Homepage, \url{http://www.springer.com/lncs}, last accessed 2023/10/25
% \end{thebibliography}
\end{document}